\documentclass[sigconf]{acmart}

\usepackage[listings]{tcolorbox}

\newcommand{\rqone}{How do large language models perform on program comprehension questions that have been generated from code created by large language models?}
\newcommand{\rqtwo}{What types of errors do large language models make when answering program comprehension questions?}

\newcommand{\qlc}[1]{\textsc{#1}}
\newcommand{\qlcb}[1]{\textbf{\qlc{#1}}}
\newcommand{\codeb}[1]{\textbf{\emph{#1}} \space}
\newtcolorbox{example}[1]{nobeforeafter,colbacktitle=white,coltitle=black!75!white,title={#1}}

\copyrightyear{2024}
\acmYear{2024}
\setcopyright{rightsretained}
\acmConference[ICSE-SEET '24]{46th International Conference on Software Engineering: Software Engineering Education and Training}{April 14--20, 2024}{Lisbon, Portugal}
\acmBooktitle{46th International Conference on Software Engineering: Software Engineering Education and Training (ICSE-SEET '24), April 14--20, 2024, Lisbon, Portugal}
\acmDOI{10.1145/3639474.3640058}
\acmISBN{979-8-4007-0498-7/24/04}

\begin{document}

\title[Let's Ask AI About Their Programs]{Let's Ask AI About Their Programs: Exploring ChatGPT's Answers To Program Comprehension Questions}

\author{Teemu Lehtinen}
\email{teemu.t.lehtinen@aalto.fi}
\orcid{0000-0003-4794-3818}
\affiliation{%
  \institution{Aalto University}
  \city{Espoo}
  \country{Finland}
}

\author{Charles Koutcheme}
\email{charles.koutcheme@aalto.fi}
\orcid{0000-0002-2272-2763}
\affiliation{%
  \institution{Aalto University}
  \city{Espoo}
  \country{Finland}
}

\author{Arto Hellas}
\email{arto.hellas@aalto.fi}
\orcid{0000-0001-6502-209X}
\affiliation{%
  \institution{Aalto University}
  \city{Espoo}
  \country{Finland}
}


\begin{abstract}
Recent research has explored the creation of questions from code submitted by students. These Questions about Learners' Code (QLCs) are created through program analysis, exploring execution paths, and then creating code comprehension questions from these paths and the broader code structure. Responding to the questions requires reading and tracing the code, which is known to support students' learning. At the same time, computing education researchers have witnessed the emergence of Large Language Models (LLMs) that have taken the community by storm. Researchers have demonstrated the applicability of these models especially in the introductory programming context, outlining their performance in solving introductory programming problems and their utility in creating new learning resources. In this work, we explore the capability of the state-of-the-art LLMs (GPT-3.5 and GPT-4) in answering QLCs that are generated from code that the LLMs have created. Our results show that although the state-of-the-art LLMs can create programs and trace program execution when prompted, they easily succumb to similar errors that have previously been recorded for novice programmers. These results demonstrate the fallibility of these models and perhaps dampen the expectations fueled by the recent LLM hype. At the same time, we also highlight future research possibilities such as using LLMs to mimic students as their behavior can indeed be similar for some specific tasks. 

\end{abstract}





\maketitle

\section{Introduction}
\label{sec:introduction}

The recent emergence of large language models (LLMs) has taken the computing education research (CER) community by storm. They have sparked discussions in conferences and yielded calls for efforts to shape the opportunities and challenges that AI-assisted programming education presents~\cite{macneil2023implications, becker2023programming,prather2023transformed}.  Multiple studies indicate that LLMs are already capable of solving assignments in programming courses~\cite{finnieansley2022robots,finnieansley2023myai,denny2023conversing,wermelinger2023using,savelka2023generative} and can explain code~\cite{sarsa2022automatic,macneil2023experiences,leinonen2023comparing,macneil2022generating}, and researchers have also explored their capability in answering beginner programmer's help requests and multiple-choice questions~\cite{hellas2023exploring,savelka2023generative,savelka2023thrilled}. As the use of LLMs becomes more prevalent in learning, one direction for evolving teaching would be to extend assessment towards program comprehension, which has been highlighted as a key skill in AI-assisted programming~\cite{becker2023programming,raman2022programming}.

Prior research in program comprehension and code tracing has highlighted that it is unclear to what extent students understand what they are building~\cite{kennedy2019qualitative,salac2020if}. Some students struggle with code tracing questions~\cite{lister2010naturally} and some fail to answer even simple multiple-choice questions about the programs that they have just written~\cite{lehtinen2023fragile}. But, what about large language models -- \textit{to what extent can large language models answer questions about code that they have generated?}

The underlying motivations for the question are threefold. Understanding the model's capacities in program comprehension will allow us to (1) study how well LLMs answer program comprehension questions, (2) how do the possible mistakes that LLMs make relate to mistakes that students make, and (3) gain an understanding of the performance and limits of the state-of-the-art models. To achieve this, we apply an existing approach used for testing students' program comprehension of their own programs~\cite{lehtinen2021lets}. However, instead of students, we target LLMs, prompting the LLMs to create solutions to programming exercises, and then using questions about learners' code (QLCs)~\cite{lehtinen2021lets} to create varying types of questions from the LLM-generated solutions that are then given back to the LLMs to solve. Our research questions for the present study are as follows:

\begin{description}
    \item [RQ1] \rqone
    \item [RQ2] \rqtwo
\end{description}

This article is organized as follows. Section \ref{sec:background} presents background in program comprehension and LLMs. Section \ref{sec:methods} explains our research method in replicable detail. Section \ref{sec:results} presents our results. Section \ref{sec:discussion} discusses our findings and limitations. Finally, Section \ref{sec:conclusions} concludes the work and outlines future research directions.

\section{Background}
\label{sec:background}


\subsection{Reading, Writing, and Tracing Code}

Novice programmers learn a wide range of skills, including reading and tracing code, writing code, and learning to understand and apply abstractions that are prevalent in code~\cite{xie2019theory}. The order in which these skills should be learned and are learned has received some attention within the CER community~\cite{lister2011concrete,lopez2009surely,xie2019theory}, also acknowledging their incremental nature that should be considered in teaching~\cite{xie2019theory}. Significant efforts have been invested into understanding (and improving) the code tracing ability of students~\cite{whalley2007many} -- one of the key observations seems to be that like writing code~\cite{mccracken2001multi}, tracing code can be challenging~\cite{lister2010naturally} and that improving both skills are part of a novice programmer's development~\cite{lister2011concrete}.

These skills are interrelated~\cite{lopez2008relationships} and practicing one of them can improve the performance of others~\cite{kumar2013study,kumar2015solving,vihavainen2015benefits}. As an example, providing code-tracing problems to students can lead to improved performance in code writing questions, even in the case where students already excel in code-tracing~\cite{kumar2013study,kumar2015solving}. Similarly, providing subtle hints (i.e. multiple-choice options that may invoke code tracing) for code explanation practice can improve subsequent code explanation performance~\cite{vihavainen2015benefits}. 

When contrasting code writing assessment and code tracing assessment, code writing might yield some points even when the answer is incorrect but resembles something that works~\cite{mccracken2001multi,traynor2006automated}. For code tracing questions, this may not be the case, as they can be marked as either being correct or incorrect~\cite{traynor2006automated,lister2010naturally}. Researchers have, however, also explored the types of mistakes students do~\cite{lister2010naturally} and sought to build a deeper understanding of issues and phases in learning to comprehend code~\cite{izu2019fostering,schulte2010introduction}. In the present work, when evaluating program comprehension questions, we use the Block Model from~\citet{schulte2008blockmodel}, which describes different areas and levels in students' understanding of program code.

\subsection{Questions About Code and Questions About Learners' Code (QLCs)}

Prompting students can have many benefits~\cite{sorva2015embedded}. When learning programming, questions about code can help students reason about the code in question, helping them as learners~\cite{lister2000blooming}. As coming up with meaningful questions can be challenging~\cite{lister2000blooming}, researchers have also proposed methods for automatically creating questions~\cite{traynor2005synthesis,qi2020unlimited,lehtinen2021lets}. These methods include mutating given code and creating code, and then asking questions about the code~\cite{traynor2005synthesis,qi2020unlimited}, as well as using code written by students as a starting point for the questions and then providing the questions back to the students who wrote the code~\cite{lehtinen2021lets}.

The latter approach, outlined by~\citet{lehtinen2021lets}, presents a design to automate questions that target concrete constructs or patterns in a student's program, and to pose these questions to the student who created that program. ~\citet{lehtinen2021lets} use the term QLCs (questions about learners' code) and discuss how they can target a range of program comprehension levels; from syntactical knowledge to design decisions. 
Tools and libraries that generate QLCs have been presented for Java~\cite{santos2022jask}, JavaScript~\cite{lehtinen2023towards}, and Python~\cite{lehtinen2023fragile}. They are all template-based designs that turn different template texts into questions by filling in facts that are extracted from the program's structure or its states during execution. They are deterministic in nature, although the generation introduces pseudo-randomness in the selection of question types and targeted elements.

Prior research on asking questions about learners' code highlights that novice students may successfully create functionally correct programs while still failing to answer simple questions about them~\cite{lehtinen2021students}. The students who repeatedly answer questions incorrectly are struggling more and resort to tinkering behavior when writing the related programs~\cite{lehtinen2023towards}. Among students who start their second programming course, a similar proportion as with complete novices still answer simple QLCs incorrectly and their answers imply their knowledge in programming is fragile~\cite{lehtinen2023fragile}.

\subsection{Large Language Models and Computing Education Research}

Large language models (LLMs) have been highlighted to function well in various areas related to learning programming, and their study in computing education research (CER) has boomed since the introduction of performant discipline-specific LLMs such as OpenAI Codex\footnote{\url{https://openai.com/blog/openai-codex}}. Works in the area have explored the possibility of using LLMs to solve programming exercises~\cite{finnieansley2022robots,finnieansley2023myai,denny2023conversing,wermelinger2023using,savelka2023generative} and to create educational content~\cite{sarsa2022automatic,denny2022robosourcing,macneil2023automatically,khosravi2023learnersourcing}, including producing explanations of code~\cite{sarsa2022automatic,macneil2023experiences,leinonen2023comparing,macneil2022generating} and improving programming error messages and feedback~\cite{leinonen2023using,phung2023generating}. LLMs have been found to outperform most of the students in introductory programming exercises~\cite{finnieansley2022robots} as well as in more complex data structures and algorithms exercises~\cite{finnieansley2023myai}. A key part of the use of LLMs is identifying meaningful prompts that direct the model to produce desired output~\cite{denny2023conversing,liu2023pre}.

The emergence of LLMs has also sparked discussions on their use in computer science classrooms~\cite{becker2023programming,denny2023computing,raman2022programming} and beyond~\cite{kasneci2023chatgpt,rajabi2023exploring,ma2023ai}, highlighting both opportunities and challenges~\cite{becker2023programming,denny2023computing,kasneci2023chatgpt,ma2023ai}. As an example, researchers have highlighted the possibility of new pedagogical approaches where LLMs are utilized for solving programming tasks~\cite{raman2022programming,becker2023programming,denny2023computing}, which could allow focusing on e.g. algorithmic concepts earlier~\cite{becker2023programming,denny2023computing}, as well as highlighted the possibilities of embedding LLMs into e.g. student's browsers to help students understand code in external resources such as StackOverflow~\cite{macneil2023experiences}.

In addition, researchers have explored how students work with LLMs~\cite{prather2023its} and evaluated the efficacy of LLMs as a tool for supporting learning~\cite{kazemitabaar2023studying}. While LLMs have so far been shown to be of help at least in introductory programming level tasks~\cite{kazemitabaar2023studying}, the produced outputs are by no means perfect. As an example, when using LLMs to answer student help requests, recent LLMs succeeded in answering slightly more than 50\% of the evaluated help requests, while also often including unwanted content such as model solutions~\cite{hellas2023exploring}. In this vein, researchers have also explored how LLMs could be used to answer other types of assessment tasks, including programming-related multiple-choice questions~\cite{savelka2023generative,savelka2023thrilled}. 




\section{Methods}
\label{sec:methods}

\subsection{Overview}

\begin{figure}[htbp]
\centering
\includegraphics[width=6.6cm]{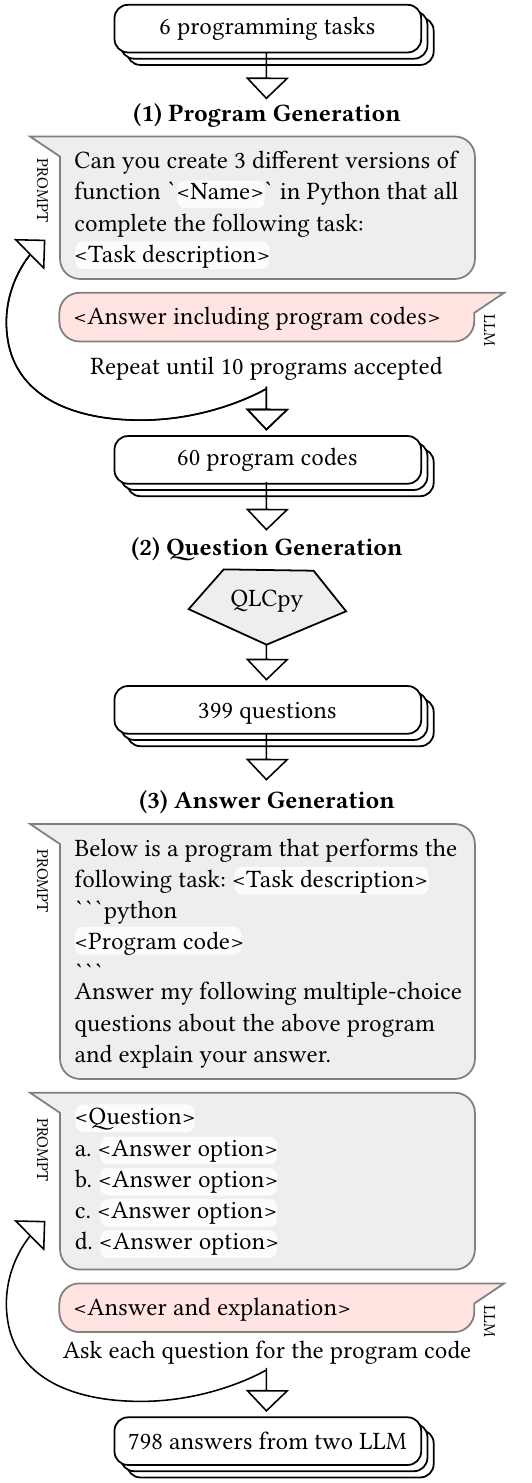}
\caption{Outline of our methodology: (1) We use LLM to generate different programs solving the same task for 6 exercises, (2) We generate QLCs for each program code, (3) We prompt LLM to answer the QLCs for each program generated earlier. This is followed by a manual analysis, discussed in Section~\ref{subsec:answer-analysis}.}
\label{fig:outline}
\end{figure}

Our overall research methodology is as follows. Given a set of programming exercise handouts, (1) we prompt an LLM to generate varying programs that solve each exercise. Then, (2) we generate multiple-choice questions out of the programs. Using the generated questions, (3) we prompt LLMs to obtain answers to them. Finally, (4) we manually analyze the answers, studying the correctness of the solutions and the errors that the LLMs make. Steps 1--3 of the methodology are outlined in more detail in Figure~\ref{fig:outline}. Next, we discuss each of the steps in our approach in more detail.

\subsection{Program Generation}
\label{sec:programs}

\subsubsection{Selected exercises.} As a starting point, we use a publicly available, large set of Python programming exercise handouts for CodeCheck\footnote{\url{https://horstmann.com/codecheck/python-questions.html}}, which were previously used in a study to evaluate LLM prompts for solving typical introductory programming tasks~\cite{denny2023conversing}. To scope the work to approximately a thousand queries for LLMs, we limited the number of included exercises to six. The selected exercises represent the range of complexity in the considered exercise set and can be solved by LLMs based on prior research~\cite{denny2023conversing}. Further, their potential solutions include different structures that are supported by the available question generators~\cite{lehtinen2021lets}. Such structures include different conditions, loops, and typical variable roles. The six selected exercises are outlined in Table~\ref{tab:pytasks}.

\begin{table}[htbp]
\centering
\caption{Selected Python programming exercises}
\label{tab:pytasks}
\begingroup
\renewcommand{\arraystretch}{1.3}
\begin{tabular}{p{17mm}p{61mm}}
\toprule
\textbf{ID}\newline {\small At CodeCheck} & Task description \\
\midrule
\textbf{T1}\newline {\small Branches, 8.}& Given three integers x, y, z, print the sum of the odd integers. \\
\textbf{T2}\newline {\small Words, 1.} & Given a string in which words are separated by spaces, return the longest word. \\
\textbf{T3}\newline {\small String\newline Operations, 1.} & Given a string s and an integer n, return a string in which each of the characters in s is repeated n times. \\
\textbf{T4}\newline {\small Averages, 4.} & Given a list of integers, return the average of all positive elements. \\
\textbf{T5}\newline {\small Double\newline Loops, 8.} & Given a list of integers, return the position of the longest subsequence of consecutive integers \mbox{a, a + 1, a + 2, \ldots} \\
\textbf{T6}\newline {\small Complex Loops, 2.} & Given a two-dimensional array of integers, remove any adjacent duplicate rows by filling the duplicates with zeroes. If more than two adjacent rows are the same, fill all but the first with zeroes. \\
\bottomrule
\end{tabular}
\endgroup
\end{table}

\subsubsection{Generating solutions.} To generate solutions to the exercises, we used OpenAI's latest GPT-3.5 model (\texttt{gpt-3.5-turbo}) with default parameters: temperature at 1, \emph{top\_p} at 1, and presence and frequency penalties at 0. We iteratively prompted the model to generate different versions of code that would solve the exercises. To query the model, we use the first prompt given in \autoref{fig:outline} where the programming exercise's task description is inserted. The reason for prompting different versions was to improve the coverage of program designs that the model is capable of producing.
Overall, we continued the prompting until we had collected ten acceptable and distinct solutions for each of the six programming exercises.
The prompt asked for three versions at a time, as we observed that asking for ten versions directly typically led to programs where the differences were very marginal, e.g. only varying few characters. With three versions at a time, the resulting body of solutions was more versatile.


\subsubsection{Rejecting incorrect, excessively complicated, or essentially identical solutions.}
When prompting the model for solutions to the programming exercises, we filtered the solutions based on the following criteria.
First, each included program had to pass online, automated functional tests of the specific CodeCheck exercise, which provided evidence that the program solved the specific exercise.
Second, we rejected programs that included module imports, lambda functions, or sophisticated generator expressions. These constructs were deemed excessively complicated for the selected introductory programming tasks as well as ignored by the currently available question generator.
Finally, the solutions had to be different from already accepted programs by either structure or identifier names. If the difference was solely on the level of identifiers, the identifier names had to have conceptual differences. For example, identically used variables named \emph{result} and \emph{out} were considered also conceptually identical while \emph{final\_string} offers more information about the variable use and was considered conceptually different from the name \emph{result}.

\subsection{Question Generation}

To automate the generation of a large number of program comprehension questions, we used the open-source QLCpy-library\footnote{\url{https://pypi.org/project/qlcpy/}} that has been previously employed to research students' comprehension of the programs they created themselves~\cite{lehtinen2023fragile}. QLCpy generates multiple-choice questions about learners' code (QLCs). The generated data includes information on the correct answers, which enables efficient assessment. Presently, the library provides eight different types of QLCs, which are outlined in \autoref{tab:qlctypes}. Each type is based on a deterministic template that is populated with facts extracted from the program to pose a unique question. QLC types require specific facts, such as the existence and position of a loop structure, that may not be available for every program. Therefore, the types of QLCs supported depend on the targeted program.

\begin{figure}[tp!]
\centering
\includegraphics[width=70mm]{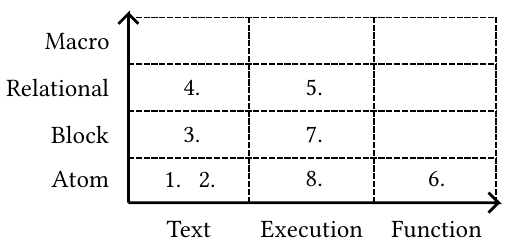}
\caption{The employed QLC types are marked with ordinal numbers and situated by the area of program comprehension they target in Schulte's Block Model~\cite{schulte2008blockmodel}}
\label{fig:blockmodel}
\end{figure}

For each LLM-generated program, we asked QLCpy to generate one QLC of every supported type.
Overall, the question generation resulted in 399 QLCs out of the 60 programs. Only 16 QLCs are of type \qlc{LinePurpose} which has the strictest requirements for the program. For each of the other seven types, the generated data has from 47 to 60 QLCs.
The employed eight QLC types outlined in Table~\ref{tab:qlctypes} target different areas of program comprehension. In terms of the Block Model represented by ~\citet{schulte2008blockmodel}, the questions work on the atom, block, and relation level, while also considering text and execution of the code, and in one case, also the function. The QLC types are mapped to the Block Model in Figure~\ref{fig:blockmodel}. 

\begin{table}[tp!]
\caption{The eight types of multiple-choice questions about learners' code (QLCs) used in this study}
\label{tab:qlctypes}
\centering
\begingroup
\renewcommand{\arraystretch}{1.3}
\begin{tabular}{lp{75mm}}
\toprule
\multicolumn{2}{l}{QLC type, template, and descriptive answer options} \\
\midrule
\textbf{1.} & \qlcb{ParameterNames} Which of the following are parameter names of the function declared on line {\it n}? \\
\textbf{2.} & \qlcb{VariableNames} Which of the following are variable names in the program? \\
\textbf{3.} & \qlcb{LoopEnd} A program loop starts on line {\it n}. Which is the last line inside it? \\
\textbf{4.} & \qlcb{VariableDeclaration} A value is [assigned to|accessed from] variable {\it variable-id} on line {\it n}. On which line is {\it variable-id} created? \\
\textbf{5.} & \qlcb{VariableRole} Which of the following best describes the role of variable {\it variable-id} that is created on line {\it n}? \\
& \qlc{Answer Options:} The variable is never accessed and could be removed; A fixed value that is not changed after created; A stepper that systematically goes through evenly spaced values; A gatherer that combines new values to itself; A holder that replaces it\'s value with the next acceptable value \\
\textbf{6.} & \qlcb{LinePurpose} Which of the following best describes the purpose of line {\it n}? \\
& \qlc{Answer Options:} Accepts new data; Guards against division by zero; Is a condition for ending program; Tells even and odd numbers apart \\
\textbf{7.} & \qlcb{LoopCount} Line {\it n} has a loop structure. How many times does the loop execute when running \mbox{\it function-id(arguments...)}? \\
\textbf{8.} & \qlcb{VariableTrace} Line {\it n} declares a variable named {\it variable-id}. Which values and in which order are assigned to the variable when running \mbox{\it function-id(arguments...)}? \\
\bottomrule
\end{tabular}
\endgroup

\end{table}

\subsection{Answer Generation}

To generate answers for the QLCs, we implemented a protocol where a user feeds an LLM a problem handout and the associated program, and then asks the LLM to respond to QLCs one by one. For the present evaluation, we used both the GPT-3.5 model (\texttt{gpt-3.5-turbo}) which is presently available for all OpenAI users and the GPT-4 model (\texttt{gpt-4}) which is in limited beta at the time of the writing of this article. Both models were queried with the default parameters: temperature at 1, \emph{top\_p} at 1, and presence and frequency penalties at 0.

The prompts are as described at the bottom of Figure~\ref{fig:outline}. We first insert a task description and an exercise solution (code) to the prompt, which is then followed by a QLC about the code. This is followed by the model producing an answer. Once the answer is recorded, we append the next QLC as a prompt to the chat. This is continued until all QLCs about the program have been asked and the answers have been recorded. In effect, each query for the OpenAI's API includes the chat session's history including the previous prompts and the model's previous responses about that program -- this history provides a type of memory that the models can use when producing outputs.

The protocol was run twice for each generated program and the associated QLCs, once for each evaluated LLM, resulting in 798 QLC answers. We note that the prompt includes program code in triple backticks which is identical to how the present GPT versions include program code in their responses. Additionally, the prompt requests to also explain the answer, which allows us to analyze the answers beyond looking at whether the answers contain one of the multiple-choice options from the provided QLCs.

\subsection{Analyzing Answers}
\label{subsec:answer-analysis}

The analysis of the answers from the LLMs is performed in two phases, answering our research questions. 

\paragraph{Answering RQ1.}  To answer the first research question, \textit{\rqone}, we randomly divided the answers from each of the models into two parts. Two researchers, each assigned one part, marked the correctness of the answers, supported by the ground-truth correctness information of the answers collected from the QLCpy-library during answer generation. Correctness was defined as the answer clearly indicating the correct option by either letter or label, without suggesting any incorrect options. For each QLC type and programming exercise, we quantitatively report the success rate for each of the two LLMs. 

\paragraph{Answering RQ2.} To answer our second research question, \textit{\rqtwo}, we study the underlying reasons for why the LLMs may fail to answer QLCs correctly. After marking the correctness of the answers as a part of answering RQ1, the two researchers met to discuss the types of errors that they observed in the data. The aim is to explore the variety of different errors and inductively generalize a coding manual to describe the observed types. 
We focused on a decisive error for each incorrect answer, which is the first detectable factor that makes the answer's \emph{explanation} incorrect while reading the answer from top to bottom.
LLMs generate text by inserting the next tokens at the end using the preceding text as input. This supports the focus on the decisive error which once generated affects the rest of the model's answer as input.
By definition, each answer can have only one decisive error and thus corresponds to a single code in the coding manual.

Using the initial coding manual the two researchers annotated one-third of the data to practice labeling and to consider whether additional codes were needed, after which they again met to discuss the data and to resolve any potential disagreements. The inter-rater reliability of this practice round (using Cohen's kappa) is included in our results. 
Following the practice and resolution of disagreements, the researchers proceeded to label the rest of the incorrect answers. In our results, we report the number of model errors on each type of coded error for each type of QLC. 

\section{Results}
\label{sec:results}

\subsection{Generated QLCs and Success Rates}

We manually analyzed the 798 QLC answers (produced from 399 QLCs given to 2 LLMs) to identify and mark answers that did not clearly suggest the correct and only the correct answer options. We identified 122 answers from GPT-3.5 that were incorrect and 47 from GPT-4 that were incorrect (61\% less than GPT-3.5). Success rates of the answers from the two LLMs per QLC type are outlined in \autoref{tab:success}.

\begin{table}[htpb!]
\centering
\caption{Number of generated QLCs (N) and the two LLMs' success rates for each QLC type}
\label{tab:success}
\begin{tabular}{lccc}
\toprule
QLC type & N & GPT-3.5 & GPT-4 \\
\midrule
1. \qlc{ParameterNames} & 60 & 95\% & 100\% \\
2. \qlc{VariableNames} & 60 & 73\% & 98\% \\
3. \qlc{LoopEnd} & 50 & 40\% & 72\% \\
4. \qlc{VariableDeclaration} & 60 & 63\% & 93\% \\
5. \qlc{VariableRole} & 47 & 83\% & 91\% \\
6. \qlc{LinePurpose} & 16 & 88\% & 100\% \\
7. \qlc{LoopCount} & 50 & 68\% & 86\% \\
8. \qlc{VariableTrace} & 56 & 55\% & 70\% \\
Over all types & 399 & 69\% & 88\% \\
\bottomrule
\end{tabular}
\end{table}

\begin{figure}[htpb!]
\centering
\includegraphics[trim=0 3mm 0 0]{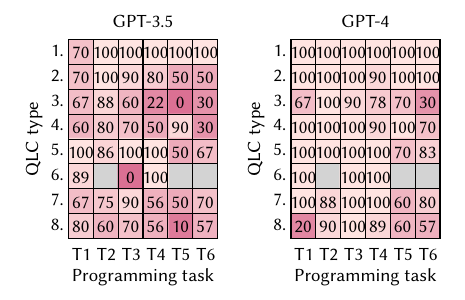}
\caption{Success rates by QLC type and programming task for the two LLMs \textnormal{--- Darker red marks lower success.}}
\label{fig:success}
\end{figure}

The success rates range from 95\% to 40\% for GPT-3.5 and from 100\% to 70\% for GPT-4 depending on the QLC type. As visible in success rates, answer quality depends on the question but it is also sensitive to the targeted program's design and features. \autoref{fig:success} presents success rates in relation to both QLC type and programming task. As the complexity of the task and resulting program code grows from T1 towards T6, success rates tend to have a decreasing trend. The state-of-the-art LLMs also have considerable difficulties for specific combinations of QLC types and program features. GPT-3.5 has success rates at or below 30\% for \qlc{LoopEnd}~(3.) on multiple tasks from T4 to T6. GPT-4 has a success rate of 20\% for \qlc{VariableTrace}~(8.) on T1, while GPT-3.5 reaches 80\% on the same QLCs. Tables are turned on tracing variables for T5 with success rates 60\% and 10\% for GPT-4 and GPT-3.5 respectively.

\subsection{Qualitative Coding of Incorrect Answers}

\begin{table}[bp!]
\centering
\caption{The coding for LLMs' incorrect answers to QLCs by the decisive error in the model's explanation}
\label{tab:coding}
\begingroup
\renewcommand{\arraystretch}{1.3}
\begin{tabular}{lp{75mm}}
\toprule
\multicolumn{2}{l}{Error code and its definition} \\
\midrule
\textbf{a.} & \codeb{Illogical execution step(s) described} Model describes the program's execution so that it differs from how the program actually executes. \\
\textbf{b.} & \codeb{Line number counted incorrectly} Model describes line contents that do not match with the line number that the model or the QLC referred to. \\
\textbf{c.} & \codeb{Interpreted question differently} Model answers logically but to a somewhat different question than posed. \\
\textbf{d.} & \codeb{Incorrect answer after valid explanation} Model explains correctly and sufficiently but finally selects contradicting answer for unknown reason. \\
\textbf{e.} & \codeb{Insufficient level of analysis} Model does not explain execution or context to the level of detail required to answer. \\
\textbf{f.} & \codeb{Valid explanation after incorrect answer} Model generates incorrect answer first but continues with contradicting, correct, and sufficient explanation. \\
\textbf{g.} & \codeb{No explanation available} Model did not generate any explanation other than the incorrect answer option(s). \\
\textbf{h.} & \codeb{Misconception about code element} Model describes a named code element as something else than it actually is. \\
\textbf{i.} & \codeb{The answer is not among the options} Model generates an answer option that was not offered in the multiple-choice question. \\
\textbf{j.} & \codeb{Hallucinates to justify incorrect answer} Model generates incorrect answer first and continues with a correct explanation until it suddenly starts to hallucinate to justify its answer. \\
\bottomrule
\end{tabular}
\endgroup
\end{table}

After identifying the 169 incorrect answers (122 from GPT-3.5 and 49 from GPT-4) and analyzing our notes, we created a coding book with 10 main decisive errors. After the two researchers had both individually coded a set of 54 incorrect answers, we calculated Cohen's kappa, resulting in a coefficient of $\kappa=0.501$. Although the inter-rater agreement was only moderate, we noticed that half of the disagreements depended on how the researchers interpreted a single code's definition. The discussion resolved all conflicts and unified our interpretation. \autoref{tab:coding} presents the final codes and their definitions.

\autoref{tab:errors} presents the frequencies of the codes for the incorrect answers for the two researched LLMs. Both models struggle with the concept of line numbers (b.) for a piece of code and illogically describe steps of execution (a.) at times. While GPT-4 considerably alleviates most error types, \emph{hallucination} (j.) becomes a frequent issue. This code describes an issue where the model provides an incorrect answer and then starts with a logical explanation, but steers towards stating falsehoods, making the issue different from illogically describing execution steps (a.).

\begin{table}[hbp!]
\centering
\caption{Number of incorrect answers from the two LLMs for each error code}
\label{tab:errors}
\begin{tabular}{lcc}
\toprule
Error code & GPT-3.5 & GPT-4 \\
\midrule
a. Illogical execution step(s) described & 31 & 10 \\
b. Line number counted incorrectly & 30 & 13 \\
c. Interpreted question differently & 12 & 6 \\
d. Incorrect answer after valid explanation & 9 &  \\
e. Insufficient level of analysis & 9 & 2 \\
f. Valid explanation after incorrect answer & 8 & 1 \\
g. No explanation available & 8 &  \\
h. Misconception about code element & 6 &  \\
i. Answer is not among the options & 5 &  \\
j. Hallucinates to justify incorrect answer & 3 & 14 \\
\bottomrule
\end{tabular}
\end{table}

\begin{figure}[hbp!]
\centering
\includegraphics[trim=0 3mm 0 0]{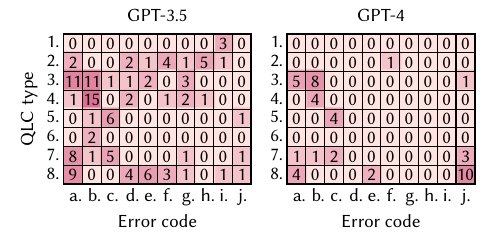}
\caption{Number of incorrect answers by QLC type and error code for the two LLMs \textnormal{--- Darker red marks higher frequency of the error code.}}
\label{fig:errors}
\end{figure}

\autoref{fig:errors} presents how error codes are distributed for different QLC types. Considering both GPT-3.5 and GPT-4, Illogical execution steps (a.) are as common for \qlc{LoopEnd}~(3.) as for the two last QLC types, \qlc{LoopCount}~(7.) and \qlc{VariableTrace}~(8.), which actually target execution of the program. Questions asking for a line number, \qlc{LoopEnd}~(3.) and \qlc{VariableDeclaration}~(4.), are plagued with issues about counting lines (b.) regardless of the LLM used. Both of the two LLMs are also struggling with the QLCs about \qlc{VariableRole}~(5.) where the majority of the errors involve interpretation of the question (c.).

While GPT-4 largely clears the issues that GPT-3.5 had at a lower frequency it still makes occasional errors, such as incorrectly interpreting (c.) what constitutes a loop's execution in \qlc{LoopCount}~(7.) and depending on insufficient level of analysis (e.) for \qlc{VariableTrace}~(8.). In contrast, GPT-3.5 has a variety of different issues (d.--i.) for QLC types targeting variables (2., 4., 8.) and loops (3., 7.).
Hallucinating an explanation to justify an incorrect answer (j.) is frequent for GPT-4. This occurs in answers about 10 programs for \qlc{VariableTrace}~(8.), 3 programs for \qlc{LoopCount}~(7.), and once for \qlc{LoopEnd}~(3.), while GPT-3.5 hallucinates 3 times in total.

\section{Discussion}
\label{sec:discussion}

\subsection{On The Performance of LLMs}

Neither of the two LLMs managed to answer all QLCs perfectly, even though they targeted programs that were generated by an LLM. GPT-4 performs considerably better than GPT-3.5 for each considered QLC type. As outlined in \autoref{tab:success} and \autoref{tab:errors}, GPT-4 provides improved performance over GPT-3.5, reducing both the number of incorrect answers and the types of errors which occur while answering. However, in the relatively small subset of programs and possible QLCs about them, as presented in \autoref{fig:success}, we discovered two combinations of programming task and QLC type where GPT-4 answered significantly less than half of the questions correctly. 

A previous study tested an earlier LLM version, specifically \texttt{text-davinci-003}, to answer multiple-choice questions on Python programming courses and report 49\% and above success rates for the first course~\cite{savelka2023generative}. Our completely different question set targets similar knowledge in Python where GPT-3.5 and GPT-4 have success rates at 40\% and 70\% or above respectively. While the studies are not comparable, the performance of GPT-4 over GPT-3.5 may work as a rough estimate of potential improvement over other earlier models. A recently published preprint from the authors of~\cite{savelka2023generative} also highlights the performance improvements of GPT-4 over GPT-3.5~\cite{savelka2023thrilled}.

Several studies have reported on students' success rates on QLCs~\cite{santos2022jask,lehtinen2023towards,lehtinen2023fragile}, where the QLCs were based on programs written by students.
The tested student populations have above 70\% success rates for QLCs that target syntactic constructs and which are similar to \qlc{ParameterNames}, \qlc{VariableNames}, \qlc{LoopEnd}, and \qlc{VariableDeclaration} in this study. When the QLCs require students to trace program's execution similarly to \qlc{LoopCount} or \qlc{VariableTrace}, the success rates are between 30\% and 50\%. Students in two different contexts answered QLCs sharing the designs of \qlc{VariableRole}~\cite{santos2022jask} and \qlc{LinePurpose}~\cite{lehtinen2023fragile} with 80\% and over 90\% success rates respectively. Although these studies on students differ from our study in the programming tasks, programming languages, and exact wording of the QLCs, students' success rates resemble those of LLMs in our study. Our results in \autoref{tab:success} suggest that GPT-3.5 may fall behind most students on QLCs that ask for line numbers and both GPT-3.5 and GPT-4 perform better than an average student on QLCs that require tracing.

Figures \ref{fig:success} and \ref{fig:errors} illustrate how both number of incorrect answers and frequency of different errors is dependent on the targeted program code as well as the devised question. Therefore, we qualitatively research different types of errors the models made and discuss our observations in the next subsections.



\subsection{LLMs and Students Make Similar Mistakes}


Generally, LLMs are good at explaining code, as highlighted by previous studies~\cite{sarsa2022automatic}, but similarly to students, they sometimes struggle to trace the program's execution at fine-grained details, especially when the code becomes more complex. We hypothesize that this trouble in tracing fine-grained code execution details led to two frequent error types. In addition to mistaking an illogical execution step (a.), the model could stay broad in the justification of its answer and produce an insufficient level of analysis (e.). As an example of insufficient analysis, GPT-3.5 created an incorrect answer that correctly explained a variable's initial value and the logic to update it. However, it did not describe execution with the given input---a requirement to justify an answer.

Illogical answers range from small errors in one step of the program to largely nonsensical explanations. For example, a \qlc{VariableTrace} question asked for values assigned to a variable that acted as a most-wanted holder for a list index. Starting from the variable's initialization to zero, GPT-4 traced the program's execution correctly except for a re-assignment of 0 on first iteration over the list, which did not change the program's logic or output. While 10 illogical explanations appeared for GPT-4, the issue is more prominent with GPT-3.5. Although we did not quantitatively report on the \emph{extent} to which the justifications are incorrect, we noticed that the model, on top of answering more questions incorrectly, has a tendency to make more errors within every single explanation compared to GPT-4. 

Beyond the model's capabilities to follow code's execution in detail, GPT-3.5 also exhibits programming misconceptions (h.). One notable example is that the model struggles with distinguishing built-in functions from declared variables. In \autoref{ex:misconception}, the program uses a built-in function named \emph{sum}, which the answer describes as a variable.

\begin{figure}[htbp!]
\centering
\begin{example}{A program generated for task T4}
\begin{lstlisting}[language=Python,numbers=left,aboveskip=0mm,belowskip=2mm,breaklines,xleftmargin=5mm]
def averageAllPositiveIntegers(numbers):
    positive_numbers = [num for num in numbers if num > 0]
    if len(positive_numbers) > 0:
        return sum(positive_numbers) / len(positive_numbers)
    else:
        return 0
\end{lstlisting}
\tcbsubtitle{A question of type \qlc{VariableNames}}
Which of the following are variable names in the program?\\
a. bar\\
b. def\\
c. if\\
d. positive\_numbers\\
e. sum
\tcbsubtitle{An answer from GPT-3.5}
d. positive\_numbers and e. sum are variable names in the program.\\
\\
Variable names are names given to a value or data stored in computer memory, and are used as references to that data. In the given program, the variables "positive\_numbers" and "sum" are used to store values, and are referenced later in the code. The other options (a. bar, b. def, and c. if) are not variable names, but rather Python keywords or reserved words. "bar" is not used in the program at all. "def" is a keyword used to define functions. "if" is also a keyword used for conditional statements in Python.
\end{example}
\caption{An incorrect answer coded as h. misconception about code element: In the program, sum is a built-in function, not a variable.}
\label{ex:misconception}
\end{figure}

The models can also miss the true intention of the question; Misinterpreting the question (c.) is also a common cause of error in both models. Still, half of these misunderstandings could be due to one specific question, \qlc{VariableRole}.
It seems that at times, the models consider a variable to be a mere \emph{holder} when the QLC library analysed it to have a role of a \emph{gatherer} or a \emph{stepper}. The model may overlook the role definitions in the question, fail to compare them, or fail to select the best description as requested and stick with a more general role.
At another level, the model may use a different concept of ``execution'', indeed, GPT-3.5 often considers that a loop ``execution'' is when it logically performs the main operation of interest in the loop. Identically, GPT-4 claimed on two occasions that a loop does not execute when a conditional statement at the beginning of its body breaks out of the loop. 

As represented by our coding of the answers, both GPT-3.5 and GPT-4 typically have a broad understanding of the function's purpose and responsibilities, which are necessary to generate code explanations. The discussed common issues---poor tracing of program's execution, not addressing the details, misinterpreting the question, and having misconceptions---are typical for humans as well.

\subsection{The Behaviour of LLMs Is Sometimes Inconsistent and Incomprehensible}

LLMs also make errors that are very specific to them and are more unlikely for humans.
One principal cause of mistakes in both GPT-3.5 and GPT-4 was line numbering (b.). The problem with identifying line numbers is reported in a previous LLM study on the Codex model \cite{sarsa2022automatic}. The models often consider the start of the code to be after the function definition. However, sometimes, the line number counting is inconsistent within the model's answer. We hypothesize that this could be due to the complexity of tracking line numbers under a textual format that is used to train these models.

Beyond the inconsistencies in tracking code blocks and line numbers, the model's full justifications may not match the chosen answer. In multiple answers, the explanation from GPT-3.5 was valid but not the answer. This occurred whether the incorrect answer was given after (d.) or before (f.) the correct explanation. 
We hypothesize that this is linked to our other observation about GPT-3.5 producing answers that are mostly non-structured. The model often mixes both answers and explanations, sometimes starting with the explanation then ending with an answer, and other times answering then justifying.

On the other hand, GPT-4 almost always first gives its answer, before it proceeds to justify it. This more structured answer matches better the implicit formatting of the prompt ``answer \emph{and} explain''. Indeed, GPT-4 is better at following instructions~\cite{openai2023gpt4} and this could, in turn, help the model alleviate its predecessors' mistakes, namely, not giving any explanation (g.), choosing an incorrect answer after a valid explanation (d.), or even giving an answer which is not among the options (i).

However, as a drawback, it could explain why GPT-4 is prone to ``hallucinate'' (j.). The term \emph{hallucination} has been coined to describe language models' tendency to create answers that are unfaithful to their input~\cite{maynez2020faithfulness}. 
In our study, GPT-4 starts with an answer that becomes fixed and then tries to justify it no matter what.
In \autoref{ex:hallucination}, we illustrate a very common error made by GPT-4 where it justifies its original answer with unexpected, sudden claims.
When such explanations would be used to support students without moderation, it could be extremely nefast for students as it could lead to misconceptions and confusion.

\begin{figure}[htbp!]
\centering
\begin{example}{A program generated for task T4}
\begin{lstlisting}[language=Python,numbers=left,aboveskip=0mm,belowskip=2mm,breaklines,xleftmargin=5mm]
def averageAllPositiveIntegers(numbers):
    sum_positive = 0
    count = 0
    for num in numbers:
        if num > 0:
            sum_positive += num
            count += 1
    if count > 0:
        return sum_positive / count
    else:
        return 0
\end{lstlisting}
\tcbsubtitle{A question of type \qlc{LoopEnd}}
A program loop starts on line 4. Which is the last line inside it?\\
a. 3\\
b. 6\\
c. 7\\
d. 8
\tcbsubtitle{An answer from GPT-4}
The last line inside the loop that starts on line 4 is:\\
b. 6\\
\\
Explanation: The loop starts on line 4 with "for num in numbers:". Lines 5, 6, and their nested statements (lines 6 and 7) are inside the loop. The loop ends on line 6, which is the last line inside the loop.
\end{example}
\caption{An incorrect answer coded as j. hallucinates to justify incorrect answer: The correct answer is line 7.}
\label{ex:hallucination}
\end{figure}

\subsection{Notes on Prompting}

We make the following general observations in terms of prompting---when trying to prompt-tune. Our ad-hoc, small ablation analysis suggests that asking the model to explain each option before giving the answers, could obtain better results and alleviate cases of hallucination. 
Furthermore, we hypothesize that the model could be asked to explicitly trace execution in the prompt to get better model answers for the types of questions where they struggle. These observations support the importance of prompt design if LLMs are used to produce descriptions that aim to guide students.

Our use of one prompting technique, ``answer and explain'', is representative of using multiple-choice questions in assessments for students
and the natural way a human would ask for the answer. Our goal was not to determine the perfect prompting strategy to obtain the best kind of answers to QLCs, although this could be interesting to explore in future work. Instead, our aim was to obtain 
insights about the underlying mistakes the model would perform when using a non-optimized prompt. 

We also note that asking for explanations already improves the model answers compared to a prompt that plainly asks to ``answer''. Without asking explanations GPT-3.5 typically does explain its answers, while GPT-4 consistently generates only the option it selects to answer. The choice of prompting with explanations after the answer is under our judgement a good prompt that allows us to understand model behaviours and performance. The answer explanations could also have a pedagogical use in asking students to assess the correctness of LLMs' answers.


\subsection{Limitations}

Like all studies, our study has a set of limitations that influence the generalizability of the results. 
First of all, we evaluated two LLMs on QLCs for six programming problems. All of the problems were selected from one context, and the problems were answered using a single programming language, Python. It is likely that selecting a different problem set and different programming problems, as well as a different programming language, would have yielded different results. The same holds also for the LLMs---while our results show differences between the two LLMs, it is possible that the performance of other (and upcoming) LLMs would differ from our results. We have researched the state-of-the-art by using the latest available models, a pervasive programming language in their training data, and typical introductory programming problems.

Second, our selection of QLCpy as the tool for generating program comprehension questions also raises limitations to our study. Although QLCpy can be used to generate different types of program comprehension questions, the questions cover only some of the aspects of code comprehension, as shown in Figure~\ref{fig:blockmodel}. It is also possible that some of the phrasings created by QLCpy (and some of the phrasings in the programming problems) influence the LLM outputs, as they are used as a part of the prompt that is used to generate answers. By using questions that have been posed to students in introductory programming classes we focus on the use of LLMs in that context.

Third, as discussed in the previous subsection, prompting plays a considerable role in producing meaningful outputs (e.g.~\cite{denny2023conversing}). 
Our results focus on a single approach that is bound to influence the outcomes of this study. The challenge with creating prompts is that even slight variations in the prompts can yield different results and thus, there is a need to balance between specificity and generalizability. In our work, we opted for a more generic prompt format over optimizing prompts for each individual exercise and QLC type separately. We acknowledge that the overall average model performance would likely have been somewhat better had we resorted to building programming problems and QLC-specific prompts. However, future use cases of LLMs may similarly limit prompt optimization if such factors remain unknown before prompt construction.

Fourth, we acknowledge that the parameters used for querying the LLMs could also be optimized. Prior work suggests that lower temperatures generate the best code exaplanations~\cite{sarsa2022automatic}. In our case, the prompts include code but are foremost natural text questions. Considering our choices about programming problems, programming language, program comprehension questions, and prompting strategy, the role of query parameters is limited, and using default values is a good starting point. However, similarly to prompt engineering, it could be interesting to explore the potential effect of query parameters in future work.

Fifth, we acknowledge that our annotation procedure is not optimal. We analyzed only the decisive errors, which we defined as the first detectable factor that makes the answer's explanation incorrect, and scoped our analysis to the responses with incorrect answer options. We acknowledge that depending on how the LLMs' explanation is potentially used, all parts of the produced text regardless of the selected options or previous errors may be significant. Such a broader study remains as future work.



\section{Conclusions}
\label{sec:conclusions}

In the present work, we explored to what extent state-of-the-art large language models (LLMs) can answer questions about code that they have generated. Our study involved using an LLM to create a body of answers (program source codes) to a set of programming exercises and then creating questions for each of those codes using the QLCpy-library, which is an open-source implementation for questions about learners' code~\cite{lehtinen2021lets}. The questions targeting code generated by the LLMs were then fed back to two state-of-the-art large language models GPT-3.5 and GPT-4, prompting them to answer the questions. Finally, we manually evaluated the answers provided by the LLMs, exploring what types of mistakes the models do when answering the questions. To summarize, our research questions and the answers to them are as follows.

\begin{description}
\item[\textbf{RQ1}:] \textit{\rqone}\newline \textbf{Answer}: Overall, large language models can answer questions about code to some extent and the introduction of the newer GPT-4 model has led to improved question-answering performance. On average, GPT-3.5 had a success rate of 69\%, while GPT-4 had an average success rate of 88\%. We also identified areas of improvement in both models that relate to the structure of the targeted programs and the type of the posed questions.
\item[\textbf{RQ2}:] \textit{\rqtwo}\newline \textbf{Answer}: We observed that both GPT-3.5 and GPT-4 make similar mistakes that have been observed with students, including not reading the question, not tracing the code, or incorrectly tracing the code. We also observed that GPT-3.5 was more likely to make mistakes and to also come up with non-existing answer options, while GPT-4 was more likely to hallucinate to justify incorrect answers. In addition, we observed that line numbers seemed to cause problems for both models.
\end{description}

Next, we present few implications that our work has for future research or potential pedagogical use of LLMs.
\begin{itemize}

\item Our work highlights that the state-of-the-art LLMs are still not perfect, even at answering simple multiple-choice questions about code. At the same time, interestingly, even though GPT-3.5 and GPT-4 do not execute the code given to them which means that they should not have the direct means for producing the expected outputs nor the ability to trace code the way how we understand tracing, they still succeed at these tasks with a high accuracy.

\item Similar to ~\cite{savelka2023thrilled}, we observed performance improvements over the evolution of the models when comparing the somewhat older GPT-3.5 with the newer GPT-4. This suggests that results in prior research using LLMs could gain improvements from adopting newer models.

\item The identified similarities between answers from students and those generated by LLMs suggest a possibility that LLM-generated data could complement data from students in research (e.g.~\cite{lister2010naturally}).
As an example, researchers looking into feedback and automated assessment in programming (e.g.~\cite{keuning2018systematic}) could explore issues that are present in content produced by LLMs, even before conducting evaluations with students. 

\item Our results support the possibility of using LLMs to enhance and automate education. As an example of an activity, students could be provided natural text answers to QLCs generated by LLMs and asked to assess the correctness of the claims. Similarly, given that students would be highlighted that the answers came from LLMs, students could also be asked to identify where the model makes a mistake. This could help in learning to read and trace code, and due to the possible mistakes in the answers, also further highlight the infallibility of the models to raise awareness about the need to critically read and evaluate outputs from such models.

\end{itemize}


As a part of our future work, we are looking into using LLMs to provide extended automated explanations on why specific answer options to QLCs are correct or incorrect. A careful study is needed on how and to what extent such justifications would help learning as their quality varies and they could sometimes include incorrect claims. Such research potentially improves QLCs but can also be generalized to discussing program code with LLMs.

We highlight the possibility of exploring the interplay of reading, writing, and tracing code for large language model researchers. As prior research into reading, writing, and tracing code has highlighted that these skills are interrelated~\cite{lopez2008relationships}, researchers looking into fine-tuning LLMs could explore to what extent fine-tuning of the code tracing functionality of LLMs influences the code reading and code generation performance of the models (similarly for fine-tuning the other aspects).

\section{Data Availability}
The data of this research as well as the program code to conduct our experiments is available at Zenodo\footnote{\url{https://doi.org/10.5281/zenodo.8434781}}.

\bibliographystyle{ACM-Reference-Format}
\bibliography{base,artorefs}


\end{document}